\newcommand*\xpartial[2]{%
  \mathrlap{\frac{\partial}{\partial #1^{\mathrlap{#2}}}}%
  \phantom{\frac{\partial}{\partial #1^{#2}}}%
  }
 \newcommand{\pa}{\partial}
 
\renewcommand{\a}{\alpha}
 \newcommand{\da}{{\dot\alpha}}
 
 \renewcommand{\b}{\beta}

 \newcommand{\ep}{\epsilon}
 \newcommand{\cN}{{\cal N}}
 \newcommand{\cO}{{\cal O}}
 \newcommand{\cM}{{\cal M}}
 \newcommand{\cZ}{{\cal Z}} 
 
\newcommand{\cT}{{\cal T}}
\newcommand{\cI}{{\cal I}}

 \newcommand{\p}[1]{(\ref{#1})}
 \newcommand \vev [1] {\langle{#1}\rangle}

\documentclass[12pt,a4paper]{article}

\usepackage{amsmath}
\usepackage{amssymb}
\usepackage{epsfig,psfrag}
\usepackage{a4wide}
\usepackage{mathtools}

\usepackage{ulem}

\csname @addtoreset\endcsname{equation}{section}

\makeatletter
\def\timenow{\@tempcnta\time
  \@tempcntb\@tempcnta
  \divide\@tempcntb60
  \ifnum10>\@tempcntb0\fi\number\@tempcntb
  \multiply\@tempcntb60
  \advance\@tempcnta-\@tempcntb
  :\ifnum10>\@tempcnta0\fi\number\@tempcnta}
\makeatother

\begin{document}

\thispagestyle{empty}


\null\vskip-43pt \hfill
\begin{minipage}[t]{50mm}
CERN-PH-TH-2015-135 \\
DCPT-15/35 \\
HU-EP-15/29\\
HU-MATH 2015-09\\
IPhT--T15/101 \\
LAPTH-032/15 
\end{minipage}

\vskip2.2truecm
\begin{center}
\vskip 0.2truecm

 {\Large\bf Bootstrapping  correlation functions in $\cN=4$ SYM  }
\vskip 0.5truecm


\vskip 0.75truecm
{\bf  Dmitry Chicherin$^{a}$, Reza Doobary$^{b}$,  Burkhard Eden$^{c}$, Paul Heslop$^{b}$,  \\[2mm] Gregory P. Korchemsky$^{d}$,
Emery Sokatchev$^{a,e,f}$ \\
}

\vskip 0.4truecm
 $^{a}$ {\it LAPTH\,\footnote{Laboratoire d'Annecy-le-Vieux de Physique Th\'{e}orique, UMR 5108},   Universit\'{e} de Savoie, CNRS,
B.P. 110,  F-74941 Annecy-le-Vieux, France\\
  \vskip .2truecm
$^{b}$   Mathematics Department, Durham University,
Science Laboratories,
\\South Rd, Durham DH1 3LE,
United Kingdom \\
\vskip .2truecm
$^{c}$ {\it Institut f\"ur Mathematik und Physik, Humboldt-Universit\"at zu Berlin, \\ Zum gro{\ss}en Windkanal 6, 12489 Berlin}
 \\
 \vskip .2truecm
$^{d}$ Institut de Physique Th\'eorique\footnote{Unit\'e Mixte de Recherche 3681 du CNRS},  CEA Saclay,  91191 Gif-sur-Yvette, France\\
\vskip .2truecm $^{e}$ Physics Department, Theory Unit, CERN, CH -1211, Geneva 23, Switzerland \\
\vskip .2truecm $^{f}$ Institut Universitaire de France,  103, bd Saint-Michel
F-75005 Paris, France 
                       } \\
\end{center}

\vskip +.75truecm

\centerline{\bf Abstract} 
\medskip
\noindent
We describe a new approach to computing the chiral part of correlation functions of stress-tensor supermultiplets in  ${\cal N} = 4$ SYM
that relies on symmetries, analytic properties and the  structure of the OPE only. 
We demonstrate that the correlation functions are given by a linear combination of chiral $\cN=4$ 
superconformal invariants accompanied by coefficient functions  depending on the space-time coordinates only. 
We present the explicit construction of these invariants and show that the six-point correlation function 
is fixed in the Born approximation up to four constant coefficients by its symmetries.   In addition, 
the known asymptotic {structure} of the correlation function in the light-like limit
 fixes unambiguously these coefficients up to an overall normalization.  
We demonstrate that the same approach can be applied to obtain a representation for the six-point NMHV amplitude that is
free from any auxiliary gauge fixing  parameters, does not involve spurious poles  and manifests half of the dual superconformal symmetry.
  
\newpage

\section{Introduction}

In this paper we continue the study of correlation functions in maximally supersymmetric $\mathcal N=4$ Yang-Mills theory ($\mathcal N=4$ SYM). 
More precisely, we shall focus on the correlation functions of local gauge-invariant operators which are members of  the stress-tensor   supermultiplet  
\begin{align}\label{Gn}
G_n \, = \, \langle \cT(1)  \cT(2) \, \ldots \, \cT(n) \rangle \,.
\end{align}
The stress-tensor supermultiplet plays a special role in $\mathcal N=4$ SYM since it comprises all conserved currents including the stress-energy
tensor as well as the Lagrangian of theory. These operators appear as coefficients in the expansion of the supercurrent $\mathcal T$ in powers of the 
Grassmann variables.  

In virtue of $\mathcal N=4$ superconformal symmetry, the two-- and three--point correlation functions (\ref{Gn}) are protected from quantum
corrections and their expressions coincide with those in the free theory.  Starting from four points, the correlation functions  (\ref{Gn}) are
not protected and depend on the coupling constant. The conjectured integrability of planar $\mathcal N=4$ SYM theory opens the
possibility of finding the exact form of this dependence, in the planar limit at least. The four-point correlation function $G_4$ has been the subject of much attention over the years. $\mathcal N=4$ superconformal symmetry fixes $G_4$ up to a single function
of the conformal cross-ratios.
At present, $G_4$ is known in planar $\mathcal N=4$ SYM theory  at weak coupling up to
seven loops in  terms of scalar conformal integrals \cite{corAmpForward} whereas the integrated expressions have been worked out up to  three 
loops~\cite{4ptintegrated12,corAmpBack,Drummond:2013nda}. At strong coupling, $G_4$ has been computed within the AdS/CFT 
correspondence in the supergravity approximation   \cite{sugraOld}.

Computing the correlation functions (\ref{Gn}) beyond four points proves to be an extremely nontrivial task. 
The conventional approach based on Feynman diagrams in configuration space is not 
suitable for $G_n$. Indeed, the contributions of the individual diagrams to $G_n$ are, in general, gauge dependent and as
a consequence, they do not respect conformal symmetry. The symmetry is only restored in the sum of all diagrams as a result 
of nontrivial cancellations of gauge dependent terms. 
Another difficulty comes from the fact that the general expression for $G_n$ satisfying the $\mathcal N=4$ superconformal Ward identities 
is given by a linear combination of nontrivial $n-$point superconformal invariants accompanied by some functions of conformal 
cross-ratios. The number of invariants as well as their complexity grow rapidly with $n$. 

This calls for developing a more efficient method for computing the correlation function (\ref{Gn}), free of the difficulties mentioned above.
The first step in this direction has been undertaken in \cite{usTwistor}. As was shown there, $G_n$ can be computed in the chiral sector 
(for all anti-chiral Grassmann variables set to zero) after reformulating $\mathcal N=4$ SYM in twistor space. This method yields
the chiral part of the correlation function $G_n$ in the Born approximation as a sum of Feynman diagrams on twistor space that involve only propagators 
and no integration vertices. The contribution of each individual diagram has a compact and concise form but it depends of the gauge fixing parameter (reference
twistor). Most importantly, it is $\mathcal N=4$ superconformally covariant modulo a compensating transformation of the 
reference twistor. The dependence on the latter disappears in the sum of all diagrams yielding the $\mathcal N=4$ symmetry
of $G_n$.\footnote{The situation here is similar to that of the
  scattering amplitudes in planar $\mathcal N=4$ SYM computed via
  twistor space~\cite{ms}.}

The question remains however whether there exists a representation for the correlation function $G_n$ that has manifest $\mathcal N=4$ superconformal symmetry and is free of any auxiliary variables
such as the reference twistor. In this paper, we argue that such a representation exists and demonstrate this by presenting an explicit
construction of the six-point correlation function $G_6$ in the chiral sector at Born level.

The paper is organised as follows. In section \ref{secConstructing} we present an ansatz for the correlation function \p{Gn} that 
obeys all available symmetry constraints. This ansatz involves just a few arbitrary constants. In section \ref{s3} we construct the explicit expression for the six-point correlation function in the Born approximation. The remaining freedom in the ansatz is fixed by requiring the known asymptotic behavior in the light-like limit. In section \ref{s5}, we demonstrate that the same approach can be applied to finding a representation for the 
six-point NMHV amplitude without unphysical spurious poles. Section 5 contains concluding remarks.
The appendix presents a technique for extracting various components of the correlation function \p{Gn} and  includes some checks
against the available data  \cite{usTwistor,dimaEmery6}. 

\section{Symmetries of the correlation functions}\label{secConstructing}

Let us recall the properties of the stress-tensor supermultiplet. 
Its lowest component is the half-BPS scalar operator
$O_{\bf 20'}(x,y)={\rm tr}\left[\Phi^I\Phi^J\right]Y^I Y^J$ built from six real scalars $\Phi^I$ (with $I=1,\dots,6$). Here $Y^I$ is a six-dimensional complex null vector that can be parametrised  as $Y^I=(1, y_{a'}^a, y^2)$ in terms of four complex variables
$y_{a'}^a$ (with $a,a'=1,2$) and $y^2=\det \| y_{a'}^a\|$.  The operator $O_{\bf 20'}(x,y)$ is annihilated by half of the Poincar\'e
supercharges, so that the stress-tensor multiplet satisfies a half-BPS shortening condition. Equivalently,  the
supercurrent $\mathcal T$ depends on half of the Grassmann variables, 
\begin{align}\label{rho}\notag
{}& \rho^{\alpha a} = \theta^{\alpha A} u_A^{+a} =  \theta^{\alpha a} + \theta^{\alpha a'} \, y^a_{a'}
\,, 
\\[2mm]
{}& \bar \rho^{\dot \alpha}_{a'} = \bar\theta^{\dot\alpha}_{\,A} \,\bar u^A_{-a'}=   \bar \theta^{\dot \alpha}_{a'}+
y^a_{a'} \, \bar \theta^{\dot \alpha}_a \,,
\end{align}
where the harmonic variables $u_A^{+a}$ and $\bar u^A_{-a'}$ (with the composite $SU(4)$ index $A=(a,a')$) parametrise  the coset $SU(4) / (SU(2) \times SU(2) \times U(1))$ (or rather its complexification). The signs $\pm$ in the indices $+a$ and $-a'$ refer to the $U(1)$ charge of the harmonics. In what follows we shall set all  $\bar \rho^{\dot \alpha}_{a'}=0$ and consider only the chiral sector of the correlation function (\ref{Gn}). 

In the chiral sector the supercurrent $\mathcal T=\mathcal T(x,y,\rho)$ depends on the four Grassmann variables
$\rho^{\alpha a}$ (with $\alpha,a=1,2$) as well as on the bosonic coordinates $x^{\dot\alpha\alpha}$ and $y_{a'}^a$. 
The advantage of introducing $y-$variables is that the $R-$symmetry acts on them in the same way as the (complexified) conformal 
group acts on the $x$'s. The supercurrent $\mathcal T(x,y,\rho)$  transforms covariantly under the $\mathcal N=4$ superconformal algebra and has conformal weight $2$ and $R-$symmetry weight $(-2)$.

\subsection{Properties of the correlation functions}

Let us examine the restrictions imposed by $\mathcal N=4$ superconformal symmetry on the correlation function (\ref{Gn}). 
It depends on the Grassmann variables $\rho_i^{\alpha a}$ (with $i=1,\dots,n$). In virtue
of $R-$symmetry,  it 
should be invariant under the center  $\mathbb Z_4$ of $SU(4)$,
$\rho_i^{\alpha a} \to e^{2\pi k/4}\rho_i^{\alpha a}$ with integer $k$. As a consequence, the (chiral) expansion of $G_n$ runs in powers of
$\rho$'s multiple of four. The lowest component of $G_n$ is $\rho-$independent, whereas the highest component
contains the product of all Grassmann variables, $\rho_1^4 \dots \rho_n^4$ (with $\rho_i^4=\prod_{a,\alpha} \rho_i^{\alpha a}$). An additional condition on the  $\rho-$dependence
comes from the invariance of $G_n$ under the chiral supersymmetry  $Q$ and antichiral conformal supersymmetry $\bar S$,
\begin{align}\label{2a}
  \rho_i{}^{\alpha a}   \ &\rightarrow\ \hat \rho_i{}^{\alpha a} =
 \rho_i{}^{\alpha a} + \left(\epsilon^{\alpha A} +x_i{}^{\alpha \dot\alpha} \bar\xi_{\dot\alpha}^A\right) u_i{}_{A}^{+a}    \,.
\end{align}
We can use the sixteen parameters of these transformations, $\epsilon^{\alpha A}$ and $ \bar\xi_{\dot\alpha}^A$, to gauge away
the same number of Grassmann $\rho-$variables. Then, the dependence on these variables can be restored by performing a finite superconformal
transformation (\ref{2a}). In this way, choosing the gauge $\rho_{n-3}^{\alpha a}=\rho_{n-2}^{\alpha a}=\rho_{n-1}^{\alpha a}
=\rho_n^{\alpha a}=0$ we find that the top component of $G_n$ takes the form $\rho_1^4 \dots \rho_{n-4}^4$. For
generic values of $\rho_i$ the chiral part of the correlation function takes the following general form for $n\ge 4$
\begin{align}\label{G-exp}
G_n = G_{n;0} + G_{n;1} +\dots+ G_{n;n-4} \,,
\end{align}
where $G_{n;p}$ is a homogenous polynomial in $\rho_1,\dots,\rho_n$ of degree $4p$ invariant under (\ref{2a}). The remaining
components vanish due to $\mathcal N=4$ superconformal symmetry, $G_{n;p}=0$ for $n-3\le p\le n$. 

Let us summarise the known properties of the components $G_{n;p}$. 

The expansion (\ref{G-exp}) is similar to that of the on-shell scattering superamplitudes in $\mathcal N=4$ SYM. This is the
reason why, by analogy with the scattering amplitude, we shall refer to $G_{n;p}$ as the N${}^p$MHV component
of the correlation function. By construction, the MHV component $G_{n;0}$ coincides with the correlation function
of the lowest component of the stress-tensor multiplet,
\begin{align}\label{Gn0}
G_{n;0} = \vev{O_{\bf 20'}(x_1,y_1) \dots O_{\bf 20'}(x_n,y_n)}\,.
\end{align}
The N${}^p$MHV component $G_{n;p}$ depends on $n$ points in the chiral analytic superspace with coordinates $(x_i,y_i,\rho_i)$. 
The Bose symmetry of the correlation function (\ref{Gn}) implies that it is invariant under the exchange of any pair of points. In 
addition, $G_{n;p}$ should have the correct conformal and $R-$symmetry transformation properties and be
invariant under the half the $\mathcal N=4$ superconformal transformations (\ref{2a})
\begin{align}\label{inv}
Q_{\alpha A} \, G_{n;p} = \bar S_{\dot\alpha}^A \, G_{n;p} = 0\,.
\end{align}
The general solution to these relations is given by a linear combination of (nilpotent) Grassmann invariants of degree $4p$
with arbitrary coefficients. We can employ the above mentioned gauge to count the total number of such invariants 
denoted by  $N_{n,p}$.
Namely, for $\rho_{n-3}^{\alpha a}=\rho_{n-2}^{\alpha a}=\rho_{n-1}^{\alpha a}
=\rho_n^{\alpha a}=0$, it is equal to the dimension of the linear space spanned by the homogenous polynomials of degree $4p$
depending on the Grassmann variables $\rho_{i}^{\alpha a}$ with $i=1,\dots,n-4$. In particular, for the bottom and top components of
(\ref{G-exp}), there is a single invariant, $1$ and $\rho_1^4\dots \rho_{n-4}^4$, respectively, leading to $N_{n,0}=N_{n,n-4}=1$.
At the same time, it is easy to see that $N_{n,p}>1$ for $1\le p\le n-5$.

At weak coupling in $\mathcal N=4$ SYM, $G_{n;p}$ admits an expansion in powers of the coupling constant
\begin{align}
G_{n;p} = \sum_{\ell\ge 0} g^{2(\ell+p)}\, G_{n;p}^{(\ell)}\,, 
\end{align}
with $0\le p\le n-4$.
The expansion starts at order $O(g^{2p})$ and the lowest term $G_{n;p}^{(0)}$ defines the Born approximation. 
The expansion coefficients $G_{n;p}^{(\ell)}$ satisfy the recurrence relations \cite{corAmpBack,corAmpForward}
\begin{align}\label{rr}
G_{n;p}^{(\ell)} =  \int d^4x_{n+1} \, d^4 \rho_{n+1} \, G^{(\ell-1)}_{n+1;p+1}\,,
\end{align}
which follows from the Lagrangian insertion method  \cite{bonus}.
Here the integral over the Grassmann variables  on the right-hand side  projects the supercurrent at point $(n+1)$ onto its
top component which is the (on-shell chiral) Lagrangian of $\mathcal N=4$ SYM theory, $\mathcal L_{\mathcal N=4}(x)= 
\int  d^4 \rho\,  \mathcal T(x,y,\rho)$. 
 
Applying (\ref{rr}) we can obtain the $O(g^{2\ell})$ correction to the $n-$point N${}^p$MHV correlation function  
by integrating the  $O(g^{2\ell-2})$ correction to the $(n+1)-$point  N${}^{p+1}$MHV correlation function.
Relation (\ref{rr}) can be iterated allowing us to obtain $G_{n;p}^{(\ell)}$ at any loop level $\ell$ as a multiple superspace
integral of the Born-level correlation function $G_{n;p+\ell}^{(0)}$. In this way, the Born-level correlation functions define
the all-loop integrands for $G_n$. For example, in the special case of $p=n-4$, the relation (\ref{rr}), combined with the 
uniqueness of the top nilpotent invariant $N_{n,n-4}=1$, has been used in \cite{corAmpBack,corAmpForward} to compute the four-point correlation function $G_4$
up to seven loops.

In planar $\mathcal N=4$ SYM, the correlation functions $G_n$ are related to the on-shell scattering amplitudes $\mathcal A_n$ through
the conjectured duality relation \cite{corAmpBos,corAmpSusy}
\begin{equation} \label{corAmp}
{\lim } \; \frac{G_n}{G_{n;0}^{(0)}} \, = \, \bigg(\frac{ \mathcal A_n}{\mathcal A_{n}^{\rm MHV}}\bigg)^2\,,
\end{equation}
where $\mathcal A_{n}^{\rm MHV}$ is the tree-level MHV amplitude and $G_{n;0}^{(0)}$ is the connected part of  (\ref{Gn0}) in the Born approximation.
Here the limiting procedure on the left-hand side amounts to putting the operators 
at the vertices of a light-like $n-$gon, $x_{i \, i+1}^2 \, = \, 0$ (with $x_{i \, i+1} = x_i - x_{i+1}$ and $x_{i+n}\equiv x_i$)
and imposing a condition on Grassmann variables, $(\theta_{i}-\theta_{i+1})^{\alpha A} (x_{i,i+1})_{\alpha\dot\alpha}=0$.
The exact identification between the coordinates of $G_n$ in the analytic superspace and the supermomenta of  $\mathcal A_n$ can be
found in \cite{corAmpSusy}.   
The duality (\ref{corAmp}) can be used   to learn about amplitudes from the knowledge of correlation functions 
 \cite{corAmpForward}. In particular, the predictions for the four-dimensional part of the amplitude integrands elaborated from  (\ref{corAmp}) using available results for the correlation functions exactly agree with the results of the recursive all-loops procedure of \cite{nima}. 

\subsection{Chiral $\mathcal N=4$ superconformal invariants}

As was explained in the previous subsection, the general expression for the correlation function $G_{n;p}$ is given by a linear
combination of chiral $\mathcal N=4$ superconformal invariants $\mathcal I_{n;p}$ accompanied by $\rho-$independent 
coefficient functions $f_{n;p}$
\begin{align}\label{sum-I}
G_{n;p} = \sum_{i}\,  \mathcal I_{n;p,i} (x,y,\rho)\, f_{n;p,i} (x,y) \,,
\end{align}
where $\mathcal I_{n;p,i}$ are functions of the $n$ points in analytic superspace invariant under (\ref{2a}) and satisfying (\ref{inv}).
Here the non-negative integer $0\le p\le n-4$
defines the Grassmann degree of the invariant and the index $i=1,\dots,N_{n,p}$  labels the different solutions to (\ref{inv}).

Taking into account that the generators $Q$ and $\bar S$ form an abelian algebra,
$\{Q_{\alpha A}, \bar S_{\dot\alpha}^B\}=0$, we can write down the general solution to (\ref{inv}) as
\begin{align}\label{ans}
\mathcal I_{n;p} {}& = Q^8 \bar S^8 \mathcal J_{n,p+4}(x,y,\rho)\,,
\end{align}
where  the right-hand side involves the product of all generators, $Q^8=\prod_{\alpha,A} Q_{\alpha A}$ and
similarly for $\bar S^8$. Since the generators $Q$ and $\bar S$ are nilpotent, $(Q_{\alpha A})^2=0$ and $(\bar S_{\dot\alpha}^A)^2=0$,
the ansatz (\ref{ans}) satisfies (\ref{inv}) for an arbitrary function $\mathcal J_{n,p+4}(x,y,\rho)$. 
Using a Grassmann integral representation for $Q^8\bar S^8$ we can rewrite (\ref{ans}) as
\begin{align}\notag\label{ans-int}
\mathcal I_{n;p} {}& = \int d^8\epsilon\, d^8\bar\xi \, {\rm e}^{\epsilon \cdot Q + \bar\xi\cdot \bar S} \mathcal J_{n,p+4}(x,y,\rho)
\\
{}& = \int d^{16}\Xi \, \mathcal J_{n;p+4}(x,y,\hat\rho)
\,,
\end{align}
where $\hat\rho_i = {\rm e}^{\epsilon \cdot Q + \bar\xi\cdot \bar S}\rho_i $ is given by (\ref{2a}) and $\Xi=(\ep,\bar\xi)$ denotes the 16 odd parameters.
By definition, $\mathcal I_{n;p}$ is a homogenous polynomial in $\rho$ of degree $4p$. Then, it follows from (\ref{ans-int}) that
$\mathcal J_{n;p+4}(x,y,\rho)$ should have the same property but its degree of homogeneity equals $4(p+4)$. 

Let us examine (\ref{ans-int}) for different Grassmann degrees $0\le p\le n-4$.

\subsubsection{Top invariant}

We start with $p=n-4$ corresponding to the top invariant $\mathcal I_{n;n-4}$. According to (\ref{ans-int}),
$\mathcal I_{n;n-4}$  is related to the function $\mathcal J_{n;n}(x,y,\rho)$. Since
$\mathcal J_{n;n}(x,y,\rho)$ depends on $n$ points and has Grassmann degree $4n$, it should necessarily involve the product of all $\rho$
variables
\begin{align}\label{Jnn}
\mathcal J_{n;n} = \rho_1^4\dots \rho_n^4\,.
\end{align}
Obviously, this Grassmann structure can be multiplied by an arbitrary function of $x$ and $y$. In the expression for the
correlation function (\ref{sum-I}) it can be absorbed into the coefficient function $f_{n;n-4}(x,y)$
\begin{align}\label{max}
G_{n;n-4} =  f_{n}(x,y) \int d^{16}\Xi \, \hat\rho_1^4\dots \hat\rho_n^4\,,
\end{align}
with $\hat\rho_i$ given by (\ref{2a}) and $f_n\equiv f_{n;n-4}$. The fact that the expression on the right-hand side contains a single term is in agreement with the 
uniqueness of the $\mathcal N=4$ superconformal invariant for $p=n-4$, $N_{n,n-4}=1$. 

Let us verify the conformal and $R-$symmetry properties of (\ref{max}). These transformations can be realised  as combinations of translations and inversions of the $x-$ and $y-$ coordinates.
In particular, under  inversion $I[x_i]=x_i^{-1}$ and $I[y_i]=y_i^{-1}$ the correlation function should acquire the 
weight $\prod_i (x_i^2)^2 (y_i^2)^{-2}$, which corresponds to   conformal weight $2$ and  $R-$charge $(-2)$ at each point.
The corresponding transformations of the  analytic superspace coordinates are
\begin{align}\label{I}
I[\rho_i{}^{\alpha a}]= \rho_i{}^{\alpha a}(x_i^{-1})_\alpha^{\dot\alpha}(y_i^{-1})^{a'}_a\,,\qquad\qquad
I[u_i{}_{A}^{+a}]= u_i{}_{A}^{+a}(y_i^{-1})^{a'}_a\,.
\end{align}
Applying inversion to (\ref{max}), we find that the $\Xi-$integral produces the weight $\prod_i (x_i^2 y_i^2)^{-2}$. Then, in 
order to reproduce the correct transformation properties of the correlation function,  the coefficient function has to satisfy
\begin{align}\label{inv-f}
I[f_{n}(x,y)] = \prod_i (x_i^2)^4 \, f_{n}(x,y)\,.
\end{align}
It follows from this relation that $f_{n}(x,y)$ is $y-$independent since otherwise it would depend on the cross-ratios $y_{ij}^2 y_{kl}^2/(y_{ik}^2 y_{jl}^2)$ yielding singularities of the correlation function for $y_{ik}^2\to 0$. This contradicts the polynomial nature of the finite-dimensional representations of $SU(4)$.  Then, the crossing symmetry of the
correlation function implies that $f_{n}(x)$ is invariant under permutations of the $n$ points, $x_i \leftrightarrow x_j$.

The coefficient function
depends on the coupling constant. Substitution of (\ref{max}) into (\ref{rr}) yields a recurrence relations that allows us to
obtain an integral representation for $f_{n}(x)$ to any loop order in terms of Born level coefficient functions $f^{(0)}_{k}(x)$ 
for $k>n$
\begin{align}
f_n^{(\ell)}(x) = \int d^4 x_5 \dots d^4 x_{5+\ell} \, f^{(0)}_{n+\ell}(x)\,.
\end{align}
The Born level coefficient functions $f^{(0)}_{k}(x)$ are totally symmetric rational functions of $x_1,\dots,x_n$ satisfying 
(\ref{inv-f}) and having only simple poles in the limit $x_{ij}^2\to 0$.\footnote{The latter property follows from the operator 
product expansion of the supercurrents.} As was demonstrated in \cite{corAmpBack,corAmpForward}, these properties alone fix the coefficient functions $f^{(0)}_{n}(x)$
up to an overall normalization constant, e.g. for $n=5, 6$ we have
\begin{align}\notag\label{f6}
{}& f^{(0)}_{5}=\frac1{\prod_{1\leq i<j\leq 5}x_{ij}^2}\ ,
\\
{}& f^{(0)}_{6} = \frac{x_{12}^2x_{34}^2 x_{56}^2  } {48\prod_{1\leq
      i<j\leq 6}x_{ij}^2} \ +\ S_6
    \text{ permutations}\,.
\end{align}
The explicit expressions for $f^{(0)}_{n}(x)$ in planar $\mathcal N=4$ SYM up to $n=11$ can be found in~\cite{corAmpBack,corAmpForward}.

\subsubsection{Next-to-top invariants}
 \label{sec:k=n-5:-next}

Now we consider the correlation function (\ref{sum-I}) for $p=n-5$. It involves next-to-top invariants $\mathcal I_{n;n-5}$ which are
related through (\ref{ans-int}) to the homogenous polynomials $\mathcal J_{n;n-1}$ of Grassmann degree $4(n-1)$. Compared with
the analogous polynomial of  maximal degree (\ref{Jnn}), we have to remove four factors of $\rho$
\begin{align}\label{Jnn-1}
\mathcal J_{n;n-5}\sim  {\partial\over\partial \rho_i^{\a a}}
{\partial\over\partial \rho_j^{\b b}}
{\partial\over\partial \rho_k^{\gamma c}}
{\partial\over\partial \rho_l^{\delta d}}\mathcal J_{n;n-4}\,.
\end{align}
Note that in the previous case  the factor of $\rho_i^4$ in (\ref{Jnn}) 
carries  $R-$symmetry weight $(-2)$ at each point. In order to preserve the $R-$symmetry properties of $\mathcal J_{n;n-5}$
we need to compensate each of the four $\rho$'s removed in (\ref{Jnn-1})
with a harmonic variable $u_{i}{}^{+a}_A$. 

 In this way, we arrive at
 \begin{align}
  \label{eq:5}
  G_{n;n-5}= \sum \cI_{ijkl;\alpha\beta\gamma\delta} \times f_{ijkl}^{\alpha\beta\gamma\delta}(x)\,,
\end{align}
where $ f_{ijkl}^{\alpha\beta\gamma\delta}\equiv f_{n;n-5}$ are coefficient functions  and the invariants
$\cI_{ijkl;\alpha\beta\gamma\delta}\equiv \mathcal I_{n;n-5}$ are given by
  \begin{align}\label{Rr}
    \cI_{ijkl ; \a \b \gamma \delta} = &  \ep^{ABCD} u_{i}{}_A^a u_{j}{}_B^b u_{k}{}_C^c u_{l}{}_D^d \int d^{16}\Xi \,  
     {\partial\over\partial\hat\rho_i^{\a a}}
{\partial\over\partial\hat\rho_j^{\b b}}
{\partial\over\partial\hat\rho_k^{\gamma c}}
{\partial\over\partial\hat\rho_l^{\delta d}}\left(
\hat\rho_1^4 \dots \hat\rho_n^4\right).
\end{align}
The sum in (\ref{eq:5}) runs over all subsets $i,j,k,l$ of four points (not necessarily different)
out of the total $n$ points. We verify using (\ref{I}) that $\cI_{ijkl;\alpha\beta\gamma\delta}$ has $R-$symmetry weight $(-2)$
at each point. As in the previous case, this leads to independence of the coefficient function $f_{ijkl}^{\alpha\beta\gamma\delta}$ of the $y-$variables.
  
As follows from the definition (\ref{eq:5}), $\cI_{ijkl;\alpha\beta\gamma\delta}$ is invariant under the simultaneous interchange of positions  and spinor indices
 \begin{align}\label{eq:11}
   \cI_{ijkl;\alpha\beta\gamma\delta}=   \cI_{jikl;\beta\alpha\gamma\delta}= \cI_{kjil;\gamma\alpha\beta\delta}\, \quad \text {etc.}
 \end{align}
This relation implies further symmetries in the cases where the positions coincide. For example, for $i=j$ we find
that $\cI_{iikl;\alpha\beta\gamma\delta}$ is symmetric in the corresponding spinor indices $\alpha$ and $\beta$
\begin{align}\label{R-sym}
\cI_{iikl;\alpha\beta\gamma\delta}= \cI_{iikl;\beta\alpha\gamma\delta}\,.
\end{align}
Moreover, for $i=j=k$ the invariant (\ref{Rr}) vanishes due to the antisymmetry of the $u-$dependent factor
\begin{align}\label{zero}
\cI_{iiil;\alpha\beta\gamma\delta}= 0\,.
\end{align}
This relation allows us to exclude the terms with three coincident position indices  from the sum in (\ref{eq:5}).
In addition, taking into account the symmetry of the correlation function 
(\ref{eq:5}) under the exchange of any pair of points, $(x_i,y_i,\rho_i) \rightarrow
(x_{\sigma(i)},y_{\sigma(i)},\rho_{\sigma(i)})$, and making use of (\ref{eq:11}), we find  that the coefficient function has
to satisfy  
\begin{align}
  \label{eq:7}
  f_{ijkl}(x_1\dots x_n)=f_{\sigma_i\sigma_j\sigma_k\sigma_l}(x_{\sigma_1}\dots x_{\sigma_n}) \end{align}
for any permutation $\sigma$ of the $n$ indices.

The invariants (\ref{Rr}) are not independent and satisfy nontrivial superconformal Ward identities
\begin{align}\label{eq:1}
\sum_{i=1}^n  X_{iM}^{\a } \cI_{ijkl; \alpha\beta\gamma\delta}   = 0
\qquad (\text{for all } j,k,l,M,\beta,\gamma,\delta)\,,
\end{align}
where the notation was introduced for 
\begin{align}
  \label{eq:3}
  X^{\alpha}_{i\,M} =(\delta^\alpha_\lambda,x^\alpha_{i,\dot \lambda}) \,,
\end{align}
with $M=(\lambda,\dot \lambda)$ being a composite index and $\alpha,\lambda,\dot\lambda=1,2$.
%
%
%
To prove these identities we rewrite (\ref{2a}) as $\hat\rho_i{}^{\alpha a}   = \rho_i{}^{\alpha a} +
X_i{}^{\alpha}_{M} \, \Xi^{MA} \, 
  u_i{}_{A}^{+a} $, so that
\begin{align}
  {\partial \over \partial \Xi^{MA}} = \sum_{i=1}^n X_{iM}^{\a } u_{i}{}_A^{+a}{\partial
    \over \partial \hat \rho_i^{\alpha a} }\ .
\end{align}
Then, we use   the definition \eqref{Rr} to get 
\begin{align}\label{id}\notag
  \sum_{i=1}^n  X_{iM}^{\a } \cI_{ijkl;  \alpha\beta\gamma\delta} {}& =
  \int d^{16}\Xi   \sum_{i=1}^n  X_{iM}^{\a } u_{i}{}_A^{+a} {\partial
    \over \partial \hat \rho_i^{\alpha a} } \Big( \ep^{ABCD}
  u_{j}{}_B^b \, u_{k}{}_C^c \, u_{l}{}_D^d 
 \\ 
{}& \times {\partial\over\partial\hat\rho_j^{\b b}} {\partial\over\partial\hat\rho_k^{\gamma c}} {\partial\over\partial\hat\rho_l^{\delta d}}\prod_{m=1}^n
\hat\rho_m^4 \Big) =
\int d^{16}\Xi    {\partial \over \partial \Xi} ({\cdots })= 0\,.
\end{align}

To obtain the correlation function (\ref{eq:5}) we have to specify the  coefficient functions $f_{ijkl}^{\alpha\beta\gamma\delta}(x)$. The main
difference compared with the previous case is that these functions carry Lorentz indices and, as  a consequence,
their conformal properties become more complicated. Nevertheless, as we demonstrate in the next section for $n=6$,
the coefficient functions in (\ref{eq:5}) are uniquely determined by the symmetry properties supplemented with the 
additional conditions coming from {the OPE}. 

It is straightforward to extend the above analysis to the correlation functions (\ref{sum-I}) with $p<n-5$. To obtain the  
invariants $\mathcal I_{n;p}$, we can use (\ref{ans-int}) and define $\mathcal J_{n;p+4}$ recursively in $p$ along the same lines as (\ref{Jnn-1}).  
However, it is a nontrivial task to find a basis of linearly independent invariants and, then, to determine the corresponding 
coefficient functions $f_{n;p}$. In the next section we show how this procedure can be carried out for $n=6$.

\section{The six-point correlation function} \label{s3}

According to (\ref{G-exp}), the six-point correlation function contains three components. The lowest, MHV component $G_{6;0}$
coincides with the correlation function of half-BPS scalar operators (\ref{Gn0}) and it is given in the Born approximation by the 
product of free scalar propagators symmetrised with respect to the permutation of the $n$ points. The connected part of $G_{6;0}$ takes
the following form
\begin{align}\label{low}
G_{6;0}^{(0)} = {y_{12}^2 \over x_{12}^2}\dots {y_{61}^2 \over x_{61}^2} + \text{$S_6$ permutations}
\end{align}
where $y_{ij}^2=\frac12 (y_{ij})^a_{a'}(y_{ij})^b_{b'}\epsilon_{ab}\epsilon^{a'b'}$ and $y_{ij}=y_i-y_j$.
The highest, N${}^2$MHV component $G_{6;2}$ has Grassmann degree $8$ and is given by (\ref{max}) for $n=6$ with
$f_6(x)$ in the Born approximation defined in (\ref{f6}). 

Let us consider the remaining NMHV component $G_{6;1}$. Applying (\ref{eq:5}) and taking into account (\ref{zero}) we find
in the Born approximation
\begin{align}
G^{(0)}_{6;1}=  
\cI_{5566;\alpha\beta\gamma\delta} \,f_{5566}^{\alpha\beta\gamma\delta}(x)
+ \cI_{4566;\alpha\beta\gamma\delta}
\,f_{4566}^{\alpha\beta\gamma\delta}(x)+\cI_{3456;\alpha\beta\gamma\delta}
\,f_{3456}^{\alpha\beta\gamma\delta}(x)  + \text{$S_6$ perm.} ,
\end{align}
with the $\cI-$invariants given by (\ref{Rr}). We recall that the coefficient functions $f_{5566}$, $f_{4566}$ and $f_{3456}$
depend only on $x$'s and have  conformal transformation  properties to be specified below. Furthermore, they are allowed to have only
simple poles, $1/x_{ij}^2$,  in the limit  $x_{ij}^2\to 0$.  To make this property manifest, we use the following
representation for the coefficient functions 
\begin{align}
  \label{eq:8}
  f_{ijkl}^{\alpha \beta\gamma\delta}(x) =\frac{p_{ijkl}^{\alpha
      \beta\gamma\delta}(x)}{ \prod_{1\le i<j\le 6}x_{ij}^2 }\,,
\end{align}
with $p_{ijkl}(x)$ being polynomials in $x$.

In summary, we have that 
\begin{align}
  \label{eq:10}
  G^{(0)}_{6;1}= \frac{
\cI_{5566;\alpha\beta\gamma\delta} \,p_{5566}^{\alpha\beta\gamma\delta}(x)
+ \cI_{4566;\alpha\beta\gamma\delta}
\,p_{4566}^{\alpha\beta\gamma\delta}(x)+\cI_{3456;\alpha\beta\gamma\delta}
\,p_{3456}^{\alpha\beta\gamma\delta}(x)}{\prod_{1\le i<j\le 6}x_{ij}^2} + \text{$S_6$ perm.} 
\end{align}
%
Let us now analyse the most general form of the polynomials
$p_{ijkl}^{\alpha\beta\gamma\delta}(x)$. Note that due to the sum over $S_6$ permutations as well as the symmetries \eqref{eq:11} of the $\cI-$invariants, there are many equivalent expressions for the $p$'s. It is enough for us to consider just one element of the equivalence class and, in particular, we do not insist that the $p$'s have the same symmetry properties as the $\cI$'s.

We show below that the procedure outlined in the previous section leaves only
ten independent numerical coefficients in the expression on the right-hand side of (\ref{eq:10}).
Moreover, the number of independent coefficients reduces to four 
after we take into account the supersymmetry Ward identities (\ref{eq:1}).

\subsection{Coefficient functions}

We recall that the correlation function $G_{6;1}$ has conformal weight $2$ at each point. 
Since  the denominator $1/\prod
x_{ij}^2$ has weight $5$ at each point, the three terms in the numerator of (\ref{eq:10}) should have weight $(-3)$ at points $1,\dots,6$. 

As follows from the definition
(\ref{Rr}), the invariant
$\cI_{5566}$ has conformal weight $(-2)$ at points $1,2,3,4$ and
tensor  weight $(-1)$ at points $5$ and $6$.  Therefore
$p_{5566}^{\alpha\beta\gamma\delta}$ must have weight  $(-1)$ at points $1,2,3,4$ and tensor weight $(-2)$ at points $5,6$.
Furthermore, as follows from (\ref{R-sym}), $\cI_{5566}^{\alpha \beta\gamma\delta}$ is
symmetric under the interchange of the Lorentz indices $\alpha\beta$ and separately of
$\gamma\delta$. Analysing the possible polynomials satisfying these conditions, we arrive at
\begin{align}\label{a}
 p_{5566}^{\alpha\beta\gamma\delta}(x)  &= a_1 (x_{51} \tilde
x_{16})^{\alpha \gamma}   (x_{52} \tilde x_{26})^{\beta \delta} x_{35}^2
x_{46}^2\ +\ 
 a_2 (x_{51} \tilde x_{16})^{\alpha \gamma}   (x_{52} \tilde
 x_{26})^{\beta \delta} x_{34}^2 x_{56}^2 \,,
\end{align}
with $a_1$ and $a_2$ being arbitrary. These are the only two independent conformal polynomials with nonvanishing contribution to  \eqref{eq:10}. Indeed, one can show that all other possibilities either reduce to  \eqref{a} or vanish after summing
over point permutations on the right-hand side of  \eqref{eq:10}.~\footnote{
By construction, $\cI_{5566}$ is invariant under an $S_4(\{1,2,3,4\}) \times S_2(\{5,6\})$ part of the entire $S_6$ symmetry. The orbit of $S_6 / (S_4 \times S_2)$ contains all the 15 point permutations of $\cI_{5566}$ while the isotropy group transforms $p_{5566}$. For example one
could imagine replacing $x_1$ and $x_2$ in (\ref{a}) with different points. But
they cannot be replaced by $x_5$ or $x_6$ since this will vanish and
replacing by $x_3$ or $x_4$ is equivalent by the $S_4$ permutation
symmetry between points 1,2,3,4. We could also
imagine having longer chains of $x$'s e.g. $(x_{51}\tilde
x_{12}x_{23}\tilde x_{36})^{\alpha\gamma}$.  But the permutation
symmetry means that points $1,2$ appear symmetrised (these points cannot appear anywhere else in $p$ due to the conformal weight) and, due to the identity $x_{51}\tilde x_{12}x_{23}+ x_{52}\tilde x_{21}x_{13} =-
x_{12}^2 x_{53}$, this reduces to previous cases. }


Similar arguments for the second term in the numerator of  \eqref{eq:10} leave the
following five possibilities for $p_{4566}^{\alpha \beta\gamma\delta}(x)$
  \begin{align}
  p_{4566}^{\alpha \beta\gamma\delta}(x)=&  \quad \, b_1 (x_{45}\tilde x_{56})^{\alpha \gamma} (x_{53}\tilde x_{36})^{\beta \delta}    x^2_{12} x^2_{46}+ b_2  (x_{43}\tilde x_{36})^{\alpha \gamma}   (x_{52}\tilde x_{26})^{\beta \delta} x^2_{16} x^2_{45} \notag \\
  &+ b_3  (x_{43}\tilde x_{36})^{\alpha \gamma} (x_{52}\tilde x_{26})^{\beta \delta}    x^2_{15} x^2_{46}+ b_4  (x_{45}\tilde x_{56})^{\alpha \gamma} (x_{53}\tilde x_{36})^{\beta \delta}    x_{16}^2 x_{24}^2\notag\\
  &+ b_5  (x_{45}\tilde x_{56})^{\alpha \gamma} (x_{54} \tilde x_{46})^{\beta
    \delta}    x^2_{16} x^2_{23}\ .\label{b}
\end{align}
This is acted upon by an $S_3 \times S_2$ symmetry, where the $S_3$ permutes points $1,2,3$, and the $S_2$ factor exchanges $4,5$ whilst simultaneously interchanging the indices
$\alpha$,$\beta$.

Finally for the third term in the numerator of  \eqref{eq:10} we have just three possibilities for $p_{3456}^{\alpha\beta\gamma\delta}(x)$
   \begin{align}
 p_{3456}^{\alpha\beta\gamma\delta}(x) =& \quad \, c_1  (x_{36}\tilde x_{65})^{\alpha \gamma} (x_{45}\tilde x_{56})^{\beta \delta}    x_{14}^2 x_{23}^2+ c_2 (x_{32}\tilde x_{25})^{\alpha \gamma}   (x_{41}\tilde x_{16})^{\beta \delta} x_{36}^2 x_{45}^2 \notag\\
  &+ c_3 (x_{36}\tilde x_{65})^{\alpha \gamma} (x_{42}\tilde x_{26})^{\beta
    \delta}    x_{15}^2 x_{34}^2 \ .\label{c}
\end{align}
One might  consider the following additional terms
\begin{align}\notag
& (x_{32}\tilde  x_{25})^{\alpha
    \gamma}   (x_{41}\tilde  x_{16})^{\beta \delta} x_{35}^2 x_{46}^2  \, ,  
\\ \notag
& (x_{36}\tilde  x_{65})^{\alpha \gamma} (x_{42}\tilde  x_{26})^{\beta \delta}
 x_{14}^2 x_{35}^2\,,  
 \\
 &  (x_{36} \tilde x_{65})^{\alpha
    \gamma} (x_{45}\tilde  x_{56})^{\beta \delta}    x_{12}^2 x_{34}^2\,,
\end{align} 
but they do not contribute to  \eqref{eq:10}. For the first two terms this is 
 due to the antisymmetry of $(x_{3j}\tilde x_{j5})^{\alpha\gamma}$ under the exchange of points $3$ and $5$, $(x_{3j}\tilde x_{j5})^{\alpha\gamma} = - (x_{5j}\tilde x_{j3})^{\gamma\alpha}$, whereas for the last term it requires a bit
more work to see this analytically.

In summary then,  simple symmetry considerations together with the understanding of the pole structure have
allowed us to reduce the freedom in the six-point NMHV correlation function $G^{(0)}_{6;1}$ to just
ten arbitrary coefficients, $a_1,a_2$, $b_1,b_2,b_3,b_4,b_5$,
$c_1,c_2,c_3$. 
The 
expression for the correlation function is then obtained by plugging \eqref{a}, \eqref{b} and \eqref{c} into \eqref{eq:10}
leading to
\begin{align}\label{ABC}
G^{(0)}_{6;1} ={1\over \prod_{1\le i<j\le 6}x_{ij}^2 }\Big[ \sum_{i=1}^3 a_i A_i + 
\sum_{j=1}^5 b_j B_j + \sum_{k=1}^3 c_k C_k \Big]\,,
\end{align}
where we introduced a notation for the $S_6$ symmetric (super)conformal polynomials, e.g. 
\begin{align}\notag\label{AB}
{}&A_1=(x_{51} \tilde
x_{16})^{\alpha \gamma}   (x_{52} \tilde x_{26})^{\beta \delta} x_{35}^2
x_{46}^2\cI_{5566;\alpha\beta\gamma\delta} + \text{$S_6$ permutations}\,,
\\[2mm]\notag
{}&A_2=(x_{51} \tilde x_{16})^{\alpha \gamma}   (x_{52} \tilde
 x_{26})^{\beta \delta} x_{34}^2 x_{56}^2\cI_{5566;\alpha\beta\gamma\delta} + \text{$S_6$ permutations}\,,
 \\[2mm]
 {}&B_2=(x_{43}\tilde x_{36})^{\alpha \gamma}   (x_{52}\tilde x_{26})^{\beta \delta} x^2_{16} x^2_{45}\cI_{4566;\alpha\beta\gamma\delta}
 + \text{$S_6$ permutations}\,,
\end{align}
and likewise for the remaining $A$, $B$ and $C$.

\subsection{Identities}

The supersymmetry Ward identity (\ref{eq:1}) leads to nontrivial relations between the various terms in (\ref{ABC}).
Multiplying both sides of (\ref{eq:1}) by the appropriate tensor structurres and
summing over the $S_6$ permutations yields  the following three independent identities involving only $A-$ and $B-$type terms
\begin{align}
  \Big( \sum_{i=1}^6 \cI_{i566;\alpha\beta\gamma\delta}(x_{i1}\tilde x_{16})^{\alpha\gamma} (x_{52}\tilde x_{26})^{\beta\delta} x^2_{34} x^2_{56}\Big) +  S_6 \text{ perm.} \ &= \ B_1+2B_3+A_2 = 0\notag\\
  \Big( \sum_{i=1}^6 \cI_{i566;\alpha\beta\gamma\delta}(x_{i1}\tilde x_{16})^{\alpha\gamma} (x_{52}\tilde x_{26})^{\beta\delta} x^2_{35} x^2_{46}\Big)   +   S_6 \text{ perm.} \ &= \ B_4+B_2+B_3+A_1 = 0\notag\\
  \Big( \sum_{i=1}^6
    \cI_{i566;\alpha\beta\gamma\delta}(x_{i1}\tilde x_{16})^{\alpha\gamma}
    (x_{52}\tilde x_{26})^{\beta\delta} x^2_{34} x^2_{56}\Big)  +  S_6 \text{
    perm.} \ &= \ B_5+2B_4+B_1 = 0\ .\label{eq:4}
\end{align}
For example, to obtain the first relation in \eqref{eq:4}, we multiply (\ref{eq:1}) by
\begin{align}
  \label{eq:9}
  \bar X_1^{M \dot \alpha} (\tilde x_{16})_{\dot \alpha}{}^\gamma
  (x_{52}\tilde x_{26})^{\beta\delta} x^2_{34} x^2_{56}\ ,
\end{align}
with $ \bar X^{M \dot \beta} = (-x^{\alpha \dot \beta},\delta_{\dot \alpha}{}^{\dot \beta})$, and then sum over $S_6$ permutations. 
Consider the six terms separately in the sum over $i$ in the first line in \eqref{eq:4}.  For $i=1$ the expression vanishes (due to the $x_{11}=0$), for $i=2$ we get  
$$
\cI_{2566;\alpha\beta\gamma\delta}(x_{21}\tilde x_{16})^{\alpha\gamma}
(x_{52}\tilde x_{26})^{\beta\delta} x^2_{34} x^2_{56}\,,
$$ 
which on swapping the points $2
\rightarrow  5 \rightarrow 4 \rightarrow 2$ and $1\leftrightarrow 3$ gives
\begin{align}
  \cI_{5466;\alpha\beta\gamma\delta}(x_{53}\tilde x_{36})^{\alpha\gamma}
  (x_{45}\tilde x_{56})^{\beta\delta} x_{12}^2 x_{46}^2\ =\ \cI_{4566;\alpha \beta \gamma
    \delta} (x_{45}\tilde x_{56})^{\alpha \gamma} (x_{53}\tilde x_{36})^{\beta \delta}
  x_{12}^2 x_{46}^2\,.\notag
  \end{align}
This term  is
equal to $B_1$ after summing all permutations. Continuing in this way,
and comparing with the $A$'s and $B$'s defined above we
obtain the right-hand side of \eqref{eq:4}.

Identities of the form $\sum_i \cI_{i456}$ which involve $B$ and $C$ are
a little less straighforward to see. However we have obtained
analytically and verified using a computer the following identities
\begin{align}\notag
&  B_4-C_1+C_3=0\,,
\\[2mm]\notag
& B_2-B_3-C_2+C_3=0\,,
\\[2mm]
& B_1 -2 \, C_3=0\,.  \label{224}
\end{align}
Combining together \p{eq:4} and (\ref{224}), we conclude that there are six identities 
between $A$, $B$ and $C$, so that \p{ABC} contains, in fact, only $10-6=4$ independent unfixed coefficients.
Choosing $A_1$, $A_2$, $B_2$ and $B_3$ as a basis we finally obtain
\begin{equation} \label{ansRed}
G_{6;1}^{(0)} \, = \, \frac{a_1' A_1 + a_2' A_2 + b_2' B_2 + b_3' B_3}{\prod_{1 \leq i < j \leq 6} x_{ij}^2}\,,
\end{equation}
with  $a_1'$, $a_2'$, $b_2'$ and $b_3'$ being arbitrary coefficients.

\subsection{Light-like limit }
\label{sec:fixing-rema-coeff}

To fix the coefficients in \p{ansRed} we shall exploit the known asymptotic behavior of the six-point NMHV 
correlation function $G_{6;1}$ in the limit where operators become light-like separated~\cite{Heslop:2001gp} 
\footnote{We discard here the contribution of the identity operator since it corresponds to a disconnected piece of the correlation function.}
\begin{align}\label{LC-OPE}
\lim_{x_{12}^2\rightarrow 0} x_{12}^2\,\cT(1) \cT(2) = y_{12}^2\,  \sum_{\cO} P_{\cO}(x_{12},\rho_{12},y_{12})\cO(2)\ ,  
\end{align}
{where the sum runs over twist-two operators $\cO$ with the coefficient functions $P_{\cO}$ being polynomial in 
 $x_{12}$, $\rho_{12}$ and $y_{12}$.}
Inserting \p{LC-OPE} into $G_{n;k}$ we  deduce
that $  \lim_{x_{12}^2,y_{12}^2\rightarrow 0}\,( x_{12}^2 G_{n;k}) =
0$ which we can use as a further constraint on the correlation function.
Imposing this constraint on the result with four
unfixed coefficients~\eqref{ansRed} 
\begin{align}
 \lim_{x_{12}^2,y_{12}^2\rightarrow 0}  \left(a_1' A_1 + a_2' A_2 +
   b_2' B_2 + b_3' B_3 \right) = 0\ ,
\end{align}
which can be easily implemented on a computer\footnote{For example, see
  the attached {\sl Mathematica} notebook where this limit is performed on
  the $\rho_6^4$ component.}, gives
$ a_1' = - 2 \, a_2' \, ,  b_2'= - 8 \, a_2' \, , b_3' =
  0\,$.
Thus we obtain the correlation function up to a single unfixed overall
constant 
\begin{align}
 G_{6;1}^{(0)}=a_2' 
 \frac{ A_2- 2 \, A_1 -8 \, B_2 }{\prod_{1 \leq i < j \leq 6} x_{ij}^2}.
\end{align}
 
Finally the overall  constant can be fixed for example by using the
amplitude/correlator duality~\cite{corAmpSusy}
which states that in the pentagon light-like limit
  \begin{align}\label{eq:12}
  \lim_{x_{12}^2,x_{23}^2,x_{34}^2,x_{45}^2,x_{51}^2 \rightarrow \, 0}
  x_{12}^2x_{23}^2x_{34}^2x_{45}^2x_{51}^2 \times G_{6;1}\big|_{\rho_6^4}
  = {y_{12}^2y_{23}^2y_{34}^2y_{45}^2y_{51}^2}{} \times 2 M_5^{(1)}(x)\ ,
\end{align}
where $M_5^{(1)}(x)$ should match the known expression for the one-loop five-point
MHV amplitude \cite{nima} in $\mathcal N=4$ SYM with the $SU(N)$ gauge
group.  We indeed find a precise match if we set the overall constant
\begin{align}
a_2'=  - \frac{ N }{480  }\,.
\end{align}
 We finally arrive at the following result for the six-point NMHV correlation function
\begin{align}
  \label{eq:2}
 G_{6;1}^{(0)}= - \frac{ N }{480  }\frac{A_2- 2 \, A_1 -8 \, B_2 }{\prod_{1 \leq i < j \leq 6} x_{ij}^2}\,,
\end{align}
with $A_1$, $A_2$ and $B_2$ given by \p{AB}. In distinction with the results on the same class of correlation functions  obtained in \cite{usTwistor} using the twistor 
space approach, the new expression \p{eq:2} is free from auxiliary gauge fixing parameters (like a reference twistor), does not
have spurious singularities  
and is manifestly $\mathcal N=4$ superconformally invariant.  Relation \p{eq:2} is the main result of this paper.

Thus we demonstrated in this section that the six-point correlation function of the stress-tensor supermultiplet 
is fixed by its symmetry properties combined with the known structure of the OPE.  
Finally we compared  the explicit expressions for various components of \p{eq:2} with those computed using both standard Feynman diagrams as well as twistor space methods.  All components agree perfectly with those found in~\cite{usTwistor,dimaEmery6}.
 
\section{Comparison with the six-point NMHV amplitude} \label{s5}

We can use the duality relation \p{corAmp} for $n=6$ to obtain from \p{eq:2} the tree-level expression for the six-point NMHV superamplitude
\begin{align}\label{dual6}
{\lim } \; \frac{G_{6;1}^{(0)}}{G_{6;0}^{(0)}} \, = \, 2 \frac{\mathcal  A_6^{\rm NMHV}}{\mathcal A_6^{\rm MHV}} \equiv 2 R_{6}^{\rm NMHV}\,,
\end{align}
where the six-point light-like limit  is specified in Appendix~\ref{App:6pt} 
and the notation is introduced  for the ratio of the tree-level superamplitudes $R_{6}^{\rm NMHV}$. Here, the expression on the left-hand side involves the 
correlation function $G_{6;0}^{(0)}$ defined in \p{low}. In the light-like limit it can be replaced by its leading asymptotic behavior given by 
the first term on the right-hand side of \p{low}. The factor of 2 on the right-hand side comes from expanding the square in \p{corAmp}.

Notice that in the six-point light-like limit,  
\begin{align}\label{limit}
[1,\dots,n] \equiv \{x_{12}^2, \dots, x_{n-1,n}^2, x_{n1}^2\to 0\}\,,
\end{align}
for $n=6$, the cyclic $S_6-$symmetry of the correlation function is broken down to the six-point dihedral symmetry   (i.e. cyclicity, $i\to i+1$, and point reversal symmetry, $i\to 7-i$) which is a symmetry of the amplitude. 
The $\cN=4$ superconformal symmetry of the correlation function leads through the duality relation \p{dual6} to the dual $\cN=4$ superconformal symmetry of the scattering amplitudes \cite{dhks}. As a consequence, the ratio function $R_{6}^{\rm NMHV}$
can be written down in a manifestly  invariant way as a sum over the dual superconformal   invariants 
\begin{align}\label{R6}
R_{6}^{\rm NMHV} = \sum_{1\le i < j\le 6} R_{*ii+1jj+1}\,.
\end{align}
These invariants admit a simple representation if rewritten  in the momentum supertwistor space $(z_{i M},\chi_i^A)$ \cite{dhks,ms}
\begin{align}\label{R-ind}
R_{ijklm} = \frac{  \delta^4\big(\langle ijkl \rangle \chi_m  + \text{ cyclic}\big)}{\langle ijkl\rangle
  \langle jklm \rangle\langle klmi \rangle\langle lmij \rangle\langle mijk \rangle}\ ,
\end{align}
where $\vev{ijkl}=\epsilon^{MNKL} z_{iM} z_{jN} z_{kK} z_{lL}$ and the argument of the Grassmann delta-function is invariant under cyclic shift of the five indices. $R_{ijklm}$ vanishes if any two indices coincide.
The invariant $R_{*ii+1jj+1}$ in \p{R6} depends on four points and the reference supertwistor
$(z_{* M},\chi_*^A)$ denoted by an asterisk.  
Replacing in  \p{dual6}  the correlation function and the ratio function by their explicit expressions, Eq.~\p{eq:2} and \p{R6}, respectively,
we verified the duality relation \p{dual6}. 

Although the $R-$invariants make the dual superconformal invariance manifest,
they obscure the known analyticity properties of the scattering amplitudes. Namely,
each individual term in \p{R6} depends on the reference twistor and, in addition,  
has non-physical spurious poles. The dependence on the reference twistor and the spurious poles 
disappear in the sum \p{R6}, although this is far from obvious. In other words, there is a conflict between the manifest dual superconformal symmetry of the invariants
and their analytic properties.
 
It is thus instructive to give up the full $\cN=4$ dual superconformal symmetry and seek another representation of the ratio function \p{R6} that 
has no spurious poles but is invariant under half of $\cN=4$ dual superconformal symmetry,   in a direct analogy to our construction of the correlation function  in the previous section.

We first note that the chiral half of the $\cN=4$ dual superconformal symmetry acts linearly on the odd components of the momentum
supertwistors $(z_{i M},\chi_i^A)$
\begin{align}
  \chi_i^A \rightarrow \hat \chi_i^A=\chi_i^A +  z_{iM} \, \Xi^{MA}\,,
\end{align}
with the same 16 odd parameters $\Xi=(\ep,\bar\xi)$ of $Q-$ and $\bar S-$transformations as before.
In close analogy with \p{ans} and \p{ans-int}, we can rewrite \p{R-ind} in a form that is manifestly 
$Q-$ and $\bar S-$invariant
\begin{align}\notag
  R_{ijklm} 
  {}&= Q^8 \bar S^8 \left[ \frac{    \chi_i ^4 \chi_j ^4 \chi_k ^4 \chi_l ^4 \chi_m ^4}{\langle ijkl\rangle
  \langle jklm \rangle\langle klmi \rangle\langle lmij \rangle\langle mijk \rangle}\right]
  \\
  {}&= \frac{\int d^{16} \Xi \,   \hat \chi_i ^4 \hat \chi_j ^4 \hat \chi_k ^4 \hat \chi_l ^4 \hat \chi_m ^4}{\langle ijkl\rangle
  \langle jklm \rangle\langle klmi \rangle\langle lmij \rangle\langle mijk \rangle} \, .
\end{align}
It is clear that this expression has a very special form. 
 At six points we can
define the most general form of the invariant (compare with the  
invariants (\ref{Rr}) for the correlation function) 
\begin{align}\label{I-inv}
  I_{ijkl} = \epsilon^{ABCD}\int d^{16} \Xi \, \frac{\pa}{\pa
    \hat\chi_i^A}\frac{\pa}{\pa \hat\chi_j^B}\frac{\pa}{\pa
    \hat\chi_k^C}\frac{\pa}{\pa \hat\chi_l^D}(\hat \chi_1)^4(\hat
  \chi_2)^4(\hat \chi_3)^4(\hat \chi_4)^4(\hat \chi_5)^4(\hat
  \chi_6)^4 \, .
\end{align}
With this definition the six-point $R-$invariant  \p{R-ind} takes the following form  
\begin{align}
  R_{ijklm}  &= \frac{ I_{pppp}}{\langle ijkl\rangle
  \langle jklm \rangle\langle klmi \rangle\langle lmij \rangle\langle mijk \rangle} 
  \end{align}
with $(i,j,k,l,m,p)$ being a permutation of the six points. Choosing the reference supertwistor in \p{R6} to be  $(z_{* M},\chi_*^A) = (z_{6 M},\chi_6^A)$, we obtain from \p{R6}
 \cite{dhks} 
\begin{align}  \label{eq:14}
R_6^\text{NMHV}=\frac{I_{1111}}{\langle2345\rangle\langle3456\rangle\langle4562\rangle\langle5623\rangle\langle6234\rangle}
\ &+\
\frac{I_{3333}}{\langle4561\rangle\langle5612\rangle\langle6124\rangle\langle1245\rangle\langle2456\rangle}\
\notag \\
&+\
\frac{I_{5555}}{\langle6123\rangle\langle1234\rangle\langle2346\rangle\langle3461\rangle\langle4612\rangle}\ .
\end{align}
Notice that $R_6^\text{NMHV}$ only contains invariants
$I_{iiii}$ with four repeated indices rather than the general
invariant $I_{ijkl}$. The reason  is that only in this case the special invariants $R_{ijklm}$ are also invariant under the other half of dual superconformal symmetry, namely $\bar Q$ and $S$.
On the other hand, $R_6^\text{NMHV}$ should only have physical poles of the form $1/\langle i\,i{+}1\,j\,j{+}1\rangle$ whereas 
the three terms in \p{eq:14} have non-physical poles, e.g. $1/\langle4562\rangle$ and $1/\langle6234\rangle$ which cancel however
in the sum (\ref{eq:14}).

Let us try to represent the six-point NMHV ratio function in the form 
\begin{align}\label{eq:15}
R_6^{\text{NMHV}} =  \frac{\sum_{i,j,k,l=1}^6  c_{ijkl}I_{ijkl}}{\langle1234\rangle\langle2345\rangle\langle3456\rangle\langle4561\rangle\langle5612\rangle\langle6123\rangle\langle1245\rangle\langle2356\rangle\langle3461\rangle},
\end{align}
where we have explicitly written the product of all allowed physical poles in the denominator, and put an arbitrary linear combination of all invariants in the numerator. The 
coefficients $c_{ijkl}$ must be polynomial functions of the bosonic twistor variables $z_{1},\dots,z_n$ only. 
Furthermore dual conformal invariance fixes these polynomials to be a product of four 
twistor four-brackets $\vev{i_1 i_2 i_3 i_4}$ with fixed homogeneity in the $z-$variables. Namely, $c_{ijkl}$ should have homogeneity $2$ at each point with an additional power at the points $i,j,k,l$ 
\begin{align}
  c_{ijkl}( \lambda_1 z_1,\dots, \lambda_6z_6) =  (\lambda_1 \dots \lambda_6)^2\lambda_i \lambda_j \lambda_k \lambda_l\,c_{ijkl}(   z_1,\dots,  z_6)  \, .
\end{align}
These properties preclude the possibility of having coefficients with three or more repeated indices, $c_{iiii}=c_{iiij}=0$ since they would necessarily include twistor four-brackets  with coinciding indices  and, hence, vanish by antisymmetry, $\vev{i_1 i_2 i_3 i_4} =-  \vev{i_2 i_1 i_3 i_4}$. For other cases we have to list all possibilities. On top of this we impose dihedral symmetry of $R_6^{\text{NMHV}}$. 

In addition, we have to take into account the superconformal Ward identities for the invariant \p{I-inv} (just as we did for the correlator \eqref{eq:1} and \eqref{224}).
In the present case we have
\begin{align}\label{new-id}
  \sum_{i=1}^6 z_{i M} I_{ijkl}=0\,.
\end{align}
Going through the calculation, we obtain from \p{eq:15} an expression for $R_6^{\text{NMHV}}$ that involves  $14$ arbitrary coefficients
 (this should be compared with the equivalent situation for the correlation function \p{ansRed} which depends on four parameters only).  
To fix these coefficients we require that \p{eq:15} should match \p{eq:14}. This yields the representation \p{eq:15} for the ratio function
which   has physical poles only, manifest (chiral)
$\cN=$ dual superconformal symmetry and the dihedral symmetry.

We can of course obtain many different representations of $R_6^{\text{NMHV}}$, depending on the choice of basis of independent invariants 
satisfying \p{new-id}. As an additional condition, we can look for a solution in which the coefficients $c_{ijkl}$ are given by the product 
of  four twistor brackets of the form $\langle i \,i{+}1\,j\,j{+}1 \rangle$, 
thus cancelling the same number of twistor brackets in the denominator of \p{eq:15}. The resulting expression for $R_6^{\text{NMHV}}$ is then given by a sum of terms each containing the product of five physical poles.
 Even with this restriction there are a number of
different forms for the amplitude. The simplest form with these
properties is 
\begin{align}\label{last1}
R^{\text{NMHV}}_{6}=\frac{1}{2}\frac{I_{1366}}{\langle 1234\rangle
  \langle 1245\rangle  \langle 1256\rangle  \langle 2345\rangle
  \langle 3456\rangle}+\text{dihedral}_{123456}\ .
  \end{align}
where `dihedral' denote 8 other terms with permuted indices needed to ensure the invariance of $R^{\text{NMHV}}_{6}$ under the cyclic shift of indices and six point
reversal.
 
The last relation should be compared with a similar expression for $n-$point NMHV amplitude found in~\cite{hodgestrnka} using a different approach. Both  expressions are cyclically
symmetric and are given by the sum of terms each involving five physical poles only. The difference 
is however that \eqref{last1}  has manifest chiral $\cN=4$ superconformal symmetry.%
\footnote{We thank Jaroslav Trnka for a discussion of
  these issues.}

\section{Conclusions}

In this paper, we have studied the chiral sector of the correlation functions of stress-tensor supermultiplets in ${\cal N} = 4$ SYM in the analytic superspace formulation \cite{gikos,survey}. As a corrolary of the $R-$symmetry, the expansion
of the correlation functions goes in powers of the chiral Grassmann variables multiple of $4$, like the scattering superamplitudes in the theory. This similarity is explained by the relation between the two quantities in the light-like limit
\cite{corAmpBos,corAmpSusy}. 

We demonstrated that the $n-$point correlation function is given by a linear combination of chiral $\cN=4$ superconformal invariants
accompanied by coefficient functions  depending only on the bosonic coordinates. We presented an explicit construction of the chiral 
$\cN=4$ superconformal invariants and showed that the form of the coefficient functions is heavily restricted by conformal symmetry, 
the internal $R-$symmetry, point permutation invariance and the absence of higher and non-physical poles.

We discussed in detail the six-point NMHV correlation function in the Born approximation. In this case, we encounter three different
types of  $\cN=4$  invariants  and the aforementioned symmetry requirements constrain the corresponding coefficient functions up to a total of ten constant
coefficients. In addition, the six-point invariants satisfy nontrivial superconformal Ward identities leading to further redundancy.
Solving these identities we were able to eliminate six constants, leaving only four unfixed parameters. 

To determine these parameters, we examined the asymptotic behavior of the correlation function in the limit where
{any two operators become light-like separated}. In this limit, the dependence of the correlation function
on the isotopic $SU(4)$ coordinates should factor out into a universal factor. We argued that the requirement
for the general ansatz for the six-point correlation function to have this factorization property fixes unambiguously all the parameters
in the Born approximation up to an overall normalization. 

We verified that, in agreement with the conjectured correlation function/scattering
amplitude duality, the obtained result for the correlation function correctly reproduces the known expressions for the five-point one-loop
MHV and six-point tree-level NMHV amplitudes.  Finally, we have shown that the same approach can be applied to obtain a representation for the scattering amplitudes that is
free from any auxiliary parameters and does not involve spurious poles.
 
There are several directions for further development of our approach. It would be interesting to extend the above analysis to correlation 
functions at higher loops and more points. Although the complexity steeply increases with the number of loops/points, we expect that,
similarly to what happens for the scattering amplitudes
in planar $\cN=4$ SYM, the final
expressions for the correlation functions should exhibit remarkable simplicity. We recall that we restricted our consideration to the chiral sector. To restore the dependence of the 
correlation functions on the antichiral Grassmann variables, we have to revisit the construction of $n-$point superconformal invariants. 
For $n=4$ this problem has been solved in \cite{Belitsky:2014zha} whereas for $n\ge 5$ it still awaits solution.

\enlargethispage{0.3 cm}

\subsection*{Acknowledgements}

D.C. is supported by the ``Investissements d'avenir, Labex ENIGMASS'' and partially supported by the RFBR grant 14-01-00341.  B.E. is supported by DFG (``Eigene Stelle" Ed 78/4-2). R.D. acknowledges support from an STFC studentship, P.H. from the STFC Consolidated Grant ST/L000407/1. R.D., P.H. and B.E. also acknowledge support from the Marie Curie network GATIS (gatis.desy.eu) of the European Union's Seventh Framework Programme FP7/2007-2013/ under REA Grant Agreement No 317089. G.K. and E.S. acknowledge partial support by the French National Agency for Research under contract BLANCSIMI-4-2011.

\appendix

\section{The computational setup: extracting components} \label{s4}

In this Appendix we describe some technical details of the calculation of the six-point NMHV correlation function.

\subsection{Expansion of superconformal invariants}

To compute various components of the six-point NMHV correlation function \p{eq:2} we need to
expand the invariants \p{Rr} in powers of $\rho$'s. As we show below this amounts to taking the  
determinant of a  $16 \times 16$ matrix and can be easily implemented on a computer.

Introducing a compact notation for $\rho_i^{\alpha_i a_i}\equiv \rho^{I_i}$ with the composite index $\cI_i=(i,\alpha_i,a_i)$
we obtain form \p{Rr} for $n=6$ points 
\begin{align}\notag\label{R-app}
 \cI_{i_5i_6i_7i_8;\alpha_5\alpha_6\alpha_7\alpha_8} {}& =\sum_{\cI_1,\cI_2,\cI_3,\cI_4} \rho_{\cI_1}\rho_{\cI_2}\rho_{\cI_3}\rho_{\cI_4}
  \,\ep^{ABCD} u_{i_1}{}_{A}^{+a_1} u_{i_2}{}_{B}^{+a_2} u_{i_3}{}_{C}^{+a_3}u_{i_4}{}_{D}^{+a_4}
 \\
{}& \times
\int d^{16} \Xi{\xpartial{\hat\rho}{{\cI_1}}} \dots {\xpartial{\hat\rho}{{\cI_8}}}[ \prod_{m=1}^6
\hat\rho_m ^4 ]\,,
\end{align}
with $\hat \rho^{\alpha a}_i = \rho{}^{\alpha a}_i + {X_{iM}{}^{\alpha} \, \Xi^{MA}} \,u_i{}_{A}^{+a}$.
Now consider the Grassmann integral in the second line of the last equation. Since it does not depend on $\rho$'s, we
may safely replace $\hat \rho_i^{\alpha a} \rightarrow X_{iM}{}^{\alpha} \, \Xi^{MA} \, u_i{}_{A}^a$ after taking $8$ derivatives 
with respect to $\hat \rho$'s. The resulting $\Xi-$integral reduces to the determinant of a certain matrix built from $X$'s and $u$'s. 

It is convenient to introduce the auxiliary $24\times 16$ matrix
\begin{align}
    \cZ^{\cI}{}_\cM  = X_{iM}{}^{\alpha} u_i{}_{A}^a
\end{align}
 so that $X_{iM}{}^{\alpha} \, \Xi^{MA} \, u_i{}_{A}^a  = \cZ^{\cI}{}_\cM \, \Xi^\cM $,
with the composite indices $\cI=(i,\alpha,a)$ and $\cM=(M,A)$ taking $6\times 2\times 2$ and $4\times 4$ values, respectively.
Then we have 
  \begin{align}\label{eq:13}
    \int d^{16} \Xi{\xpartial{\hat\rho}{{\cI_1}}} \dots {\xpartial{\hat\rho}{{\cI_8}}}[ \prod_{m=1}^6
(\hat\rho_m)^4 ]= \big[\cZ\big]_{\cI_1\cI_2\cI_3\cI_4\cI_5\cI_6\cI_7\cI_8}\,,
  \end{align}
where by $[\cZ]_{\cI_1\cI_2\cI_3\cI_4\cI_5\cI_6\cI_7\cI_8}$ we denote the maximal $16 \times 16$ minor obtained by taking the determinant of the matrix obtained from  the $24 \times 16$ matrix $\cZ$ by removing the rows $ \cI_1,\cI_2,\dots \cI_8$.

Applying \p{R-app} and \p{eq:13} we can expand the polynomials \eqref{AB} in powers of the Grassmann variables, e.g.
\begin{align}
  A_1=\, & \sum_{\cI_1,\cI_2,\cI_3,\cI_4} \rho_{\cI_1}\rho_{\cI_2}\rho_{\cI_3}\rho_{\cI_4}\Big\{ 
  [\cZ]_{(5\alpha a)(5\beta b)(5\gamma c)(5\delta d)\cI_1\cI_2\cI_3\cI_4} \,
  \ep^{ABCD} u_{5}{}_{A}^{a} \, u_{5}{}_{B}^{b} \, u_{6}{}_{C}^{c} \, u_{6}{}_{D}^{d} \,
    \, \notag 
\\ 
&\qquad \qquad  \times  (x_{51} \tilde
x_{16})^{\alpha \gamma}   (x_{52}\tilde  x_{26})^{\beta \delta} x_{35}^2
x_{46}^2 + S_6 \ \text{permutations}\Big\},
\end{align}
where the permutations act only on the particle numbers $1,2,3,4,5,6$ (and do not act on  $\cI_1,\dots,\cI_4$). 
It is straightforward enough to implement this relation on {\sl Mathematica}. For illustration we attach a notebook to the arXiv submission of the article.

We can further simplify the calculation by making use of conformal symmetry. We recall that the polynomials \p{AB} have conformal
weight $(-3)$ at each point. Then, the conformal symmetry allows us to fix four out of six space-time coordinates $x_i^{\da\a}$. 
For example, we can put $x_1^{-1} \rightarrow 0, \, x_2$ diagonal $2 \times 2$ matrix, $x_3 \rightarrow \mathbb{I}_2, \, x_4 \rightarrow 0$
while $x_5, x_6$ are left as arbitrary $2 \times 2$ matrices. Similar choices can be also made for the $y_i$ variables. In this gauge,
the calculation of the determinant in \p{eq:13} is simplified significantly. It is then straightforward to reconstruct the fully covariant answer.

\subsection{The five-point light-like limit}

We can fix the coefficients in \p{ansRed} by examining the asymptotic behaviour of the six-point correlator in the light-like limit \p{eq:12}.

 In order to define the five-point light-like limit \p{limit}, it is convenient to make use of the variables $X^{\alpha}_{i\,M}$ introduced in \p{eq:3}.
 Then, it is easy to see that
 \begin{align}\label{sq}
x_{ij}^2 \sim  \epsilon_{\alpha\beta}\epsilon_{\gamma\delta} \epsilon^{MNKL} X_{i,M}^{\alpha}X_{i,N}^{\beta}X_{j,K}^{\gamma}X_{j,L}^{\delta}\,.
\end{align}
This relation is invariant under ocal $SL(2)$ transformations $X^{\alpha}_{i\,M}\to g^\alpha_\beta(x_i) X^{\beta}_{i\,M}$. Using this property,
we can realise the five-point light-like limit by assigning the following values of $X^{\alpha}_{i\,M}$ for $\alpha=1,2$
\begin{align}\label{5ptlim}
X^{\alpha}_{i,M}=(z_{i M},z_{i-1\, M})\,,\qquad 
X^{\alpha}_{6,M}=(z_{7 M},z_{6 M})\,,\qquad (1 \leq i \leq 5)
\end{align}
 with $z_{0,M}=z_{5,M}$. Indeed, we find from \p{sq} that for $1\le i,j\le 5$
 \begin{align}
x_{ij}^2\sim \epsilon^{MNKL} z_{i M} z_{i-1 N} z_{j K} z_{j-1 L} \equiv \vev{i,i-1,j,j-1}\,,
\end{align}
leading to $x_{i,i+1}^2\to 0$ with $x_{i6}^2\neq 0$, in agreement with \p{limit}.  
  
We expect from \p{eq:12} that the $O(\rho_6^4)$ component of the correlation function  \p{ansRed} should factorise in the kinematics 
\p{5ptlim} into the product of $y_{12}^2\dots y_{51}^2$ and an $x-$dependent function. Since $\rho_6^4$ carries the required $R-$charge at point
$6$, its coefficient can only depend on $y_1,\dots,y_5$. The latter dependence is constrained by the $R-$symmetry that acts on the $y$'s very much the same as the conformal group on the $x$'s. Examining all possible polynomials in $y_{ij}^2$ that transform covariantly under $R-$symmetry
with  weight $(-2)$ at points $1,\dots,5$, 
we find that
\begin{equation}
y_{12}^2 y_{23}^2 y_{34}^2 y_{45}^2 y_{51}^2 \, , \qquad (y_{12}^2)^2 y_{34}^2 y_{45}^2 y_{53}^2 \label{ally5s}
\end{equation}
and their eleven and nine $S_5$ permutations, respectively, are the only structures that can occur. Following  \p{eq:12} we have to impose the absence of all but the first of these.

As mentioned above, we can simplify the calculation by choosing an appropriate gauge for the $x-$ and $y-$variables.  The conformal symmetry can be used to put $z_{i N} \, = \, \delta_{iN}$ for $1 \leq i \leq 4$ in \p{5ptlim} while $z_5, z_6, z_7$ remain general. For the $y-$variables we use $R-$symmetry to choose  
\begin{equation}
y_1 \, \to \, \infty \, \mathbb{I}_2 \, , \qquad y_2 \, = \,  \left( \begin{array}{cc} y & 0 \\ 0 & \bar y \end{array} \right) \, , \qquad y_3 \, = \, \mathbb{I}_2 \, , \qquad y_4 \, = \, 0\,, \label{simplifyY}
\end{equation}
while $y_5$ remains general; $y_6$ drops out because $\rho_6^4$ saturates the $R-$charge at point 6. Going through the calculation, we obtain
the following expression for a particular $O(\rho_6^4)$ component of the correlation function \p{ansRed} in the light-like limit \p{eq:12}  
\begin{align}\notag\label{last}
  \lim_{[1,\dots,5]}\, x_{12}^2x_{23}^2  {}&  x_{34}^2x_{45}^2x_{51}^2  \times G_{6;1}\big|_{\rho_6^4} 
   ={y_{14}^2 y_{15}^2 (y_{23}^2)^2 y_{45}^2
  \over \langle 1267 \rangle \langle 2367 \rangle \langle 3467 \rangle \langle 4567 \rangle \langle 5167 \rangle} 
  \\\notag
   \times{}& 
 \Big\{ \, 8  (a_1' + 2 \, a_2' - b_3')[ \langle 1345 \rangle \langle 2345 \rangle \langle 1267 \rangle  
+   \langle 1245 \rangle \langle 1345 \rangle \langle 2367 \rangle ]
\\\notag
{}& +  2 \, (-2 \, a_1' - 4 \, a_2' + 3 \, b_3')[ \langle 1235 \rangle \langle 1245 \rangle \langle 3467 \rangle + \langle 1234 \rangle \langle 2345 \rangle \langle 1567 \rangle   ]    \\
{}&   +  2  (-8 \, a_1' + 2 \, b_2' + 3 \, b_3')  \langle 1345 \rangle \bigl(\langle 5127 \rangle \langle 2346 \rangle - \langle 5126 \rangle \langle 2347 \rangle \bigr)  
  \nonumber \\
{}&    +  2  (6 \, a_1' - 4 \, a_2' - 2 \, b_2' - b_3') \langle 1234 \rangle \langle 1235 \rangle \langle 4567 \rangle   \Big\}   +\dots
\end{align}
where $\vev{ijkl}=\epsilon^{NMKL} z_{iM} z_{jN} z_{kK} z_{lL}$ and ellipses denote terms with other $y-$structures. According to  \p{ansRed}, the coefficient in front of $y_{14}^2 y_{15}^2 (y_{23}^2)^2 y_{45}^2$ should vanish.  Putting the right-hand side of \p{last} to zero yields
\begin{equation}
a_1' \, = \, - 2 \, a_2' \, , \qquad  b_2' \, = \, - 8 \, a_2' \, , \qquad b_3' \, = \, 0\,.
\end{equation} 
We verified that this choice eliminates in fact all unwanted $y-$structures leading to
\begin{align}\notag\label{MHV5}
{}&  \lim_{[1,\dots,5]}\, x_{12}^2x_{23}^2     x_{34}^2x_{45}^2x_{51}^2  \times G_{6;1}\big|_{\rho_6^4} 
   =-960 \, a_2' \, y_{12}^2 y_{23}^2 y_{34}^2 y_{45}^2 y_{51}^2
   \\
{}& \times    \frac{
\langle 1234 \rangle \langle 2345 \rangle \langle 1567 \rangle +
\langle 1235 \rangle \langle 1245 \rangle \langle 3467 \rangle +
\langle 1345 \rangle \bigl(\langle 5127 \rangle \langle 2346 \rangle - 
   \langle 5126 \rangle \langle 2347 \rangle \bigr)}{\langle 1267 \rangle \langle 2367 \rangle \langle 3467 \rangle \langle 4567 \rangle \langle 5167 \rangle}
\end{align}
Although this is not manifest, this relation is invariant under the cyclic shifts of points $1 \ldots 5$. We verified that  \p{MHV5} agrees with the known result for the four-dimensional integrand of the one-loop five-point MHV amplitude \cite{nima} with
the normalization constant $a_2'$ given by the following expression for an $SU(N)$ gauge group
\begin{equation}
a_2' \, = \, - \frac{  N }{480 }\,.
\end{equation} 

\subsection{The six-point light-like limit}\label{App:6pt}

As another test of \p{eq:2} we can examine the asymptotic behavior of $G_{6;1}^{(0)}$ in the six-point light-like limit, Eq.~\p{limit} for $n=6$.
According to \p{corAmp}, we expect to recover in this limit the known expression for the six-particle tree-level NMHV amplitude  \cite{dhks,ms}.

The analysis goes along the same lines as in the five-point light-like limit. 
The analogue of formula (\ref{5ptlim}) is
\begin{equation}\label{X6}
X^\alpha_{iM} \, = \,  (z_{iM} \, ,  z_{i-1 \, M}) \, , \qquad 1 \leq i \leq 6 \, , \qquad z_0  \, = \,  z_6\,.
\end{equation}
The six-point scattering amplitude $A_{ 6}^{\rm NMHV}$ depends on momentum twistors $z_{i,M}$ and their supersymmetric counter-parts $\chi_i^A$. The latter are related to the Grassmann variables $\theta_i^{\alpha A}$ entering \p{rho}  in the same fashion as  \p{X6}
\begin{equation}
\theta_i^{\alpha A} \, = \, (\chi_i^A \, , \chi_{i-1}^A) \, , \qquad 1 \leq i \leq 6 \, , \qquad \chi_0 \, = \, \chi_6 \,,
\end{equation}
or equivalently
$
  \rho_i^{\alpha a} \, = \,   (\chi_i^A \, {u_i}_A^{+a}   ,  \, \chi_{i-1}^A \, {u_i}_A^{+a})
$.
Going through the calculation we found that
\begin{align}
 \lim_{[1,\dots,6]}\, x_{12}^2x_{23}^2   x_{34}^2x_{45}^2x_{56}^2x_{61}^2  \times G_{6;1}^{(0)} 
  \sim y_{12}^2 y_{23}^2 y_{34}^2 y_{45}^2 y_{56}^2y_{61}^2\, A_{6}^{\rm NMHV}\,,
\end{align}
in agreement with  \p{corAmp}. The details can be found in a {\sl Mathematica} notebook
 included with the arXiv submission of this article.

\end{document}